\documentclass[lettersize, journal]{IEEEtran} 
\usepackage{cite}
\usepackage{amsmath,amssymb}
\usepackage{amsthm}
\allowdisplaybreaks
\usepackage[ruled,vlined]{algorithm2e}
\usepackage{setspace}
\usepackage{caption}
\usepackage{subcaption}
\usepackage{tikz}
\usepackage{tkz-euclide}
\usepackage{algorithm2e}
\usepackage[long]{optidef}
\usepackage{fancyhdr}

\newtheorem{theorem}{Theorem}

\fancypagestyle{firstpage}{%
  \fancyhf{} 
  \fancyhead[C]{%
    \begin{minipage}{\textwidth}
    \centering
    \scriptsize
    This work has been submitted to the IEEE for possible publication. 
    Copyright may be transferred without notice, after which this version may no longer be accessible.
    \end{minipage}%
  }%
}

\begin{document}

\title{Dual-Function Radar-Communication Beamforming with Outage Probability Metric}
\author{\IEEEauthorblockN{Hossein Maleki,$^{1}$ \IEEEmembership{Graduate Student Member, IEEE}, Carles Diaz-Vilor,$^{1}$  \IEEEmembership{Member, IEEE}, Ali Pezeshki,$^{2}$ \IEEEmembership{Senior Member, IEEE},  Vahid Tarokh,$^{3}$ \IEEEmembership{Fellow, IEEE}, and Hamid Jafarkhani$^{1}, \IEEEmembership{Fellow, IEEE} $\thanks{This work was supported
in part by the collaborative NSF Award CNS-2229467/CNS-2229468/CNS-2229469.}}

\IEEEauthorblockA{$^{1}$Center for Pervasive Communications and Computing, University of California, Irvine, CA 92697\\
$^{2}$Colorado State University, Fort Collins, CO 80523 \\
$^{3}$Duke University, Durham, NC 27708}

}

\maketitle
\thispagestyle{firstpage} 

\begin{abstract}
    The integrated design of communication and sensing may offer a potential solution to address spectrum congestion. In this work, we develop a beamforming method for a dual-function radar-communication system, where the transmit signal is used for both radar surveillance and communication with multiple downlink users, despite imperfect channel state information (CSI). We focus on two scenarios of interest: radar-centric and communication-centric. In the radar-centric scenario, the primary goal is to optimize radar performance while attaining acceptable communication performance. To this end, we minimize a weighted sum of the mean-squared error in achieving a desired beampattern and a mean-squared cross correlation of the radar returns from directions of interest (DOI). We also seek to ensure that the probability of outage for the communication users remains below a desired threshold. In the communication-centric scenario, our main objective is to minimize the maximum probability of outage among the communication users while keeping the aforementioned radar metrics below a desired threshold. 
    Both optimization problems are stochastic and untractable. We first take advantage of central limit theorem to obtain deterministic non-convex problems and then consider relaxations of these problems in the form of semidefinite programs with rank-1 constraints. We provide numerical experiments demonstrating the effectiveness of the proposed designs.
\end{abstract}
\section{Introduction}
In recent years, spectrum sharing among different users and applications has emerged as a potential solution to increasing demand for wireless communication data \cite{fcc2024spectrum}. In particular,  spectrum sharing between radar and communication systems over the same frequency band is being extensively studied (see, e.g., \cite{liu2020joint, 9585321, mishra2019toward, chiriyath2017radar}). The joint operation of radar and communication occurs primarily through two modalities: dual-function radar-communication (DFRC) and radar-communication coexistence (RCC). DFRC systems perform both radar and communication functionalities using a shared transmitter \cite{10124714, 10027173, 10077343, 9854898}. In RCC, separate radar and communication systems share information (to various degrees) to manage the interference \cite{zheng2019radar, 10287644, 9729741, 10258053}. The design problems associated with these modalities can be divided into two main categories: radar-centric and communication-centric, depending on whether the priority is given to sensing or communication. 

In multiple-input multiple-output (MIMO) DFRC systems, a critical challenge is the design of beamforming vectors that meet the demands of both communication and radar systems. 
This task is often accomplished by formulating an optimization problem that considers the performance metrics of both radar and communication systems. 
Some common metrics for evaluating radar performance include beampattern matching error \cite{liu2021dual, hua2023optimal}, the mean-squared cross-correlation between radar returns from different angles (bearing response pattern) \cite{stoica2007probing, liu2020mimo}, the radar receiver’s signal-to-interference-plus-noise ratio (SINR) \cite{liu2022joint, qian2018joint}, and Cram\'er-Rao lower bound (CRLB) \cite{yin2022rate, kumari2019adaptive}. Similarly, typical metrics used to evaluate the performance of the communication systems are the SINR of individual users \cite{qi2022hybrid, wen2023transmit} and sum rate \cite{yuan2020bayesian, cheng2021hybrid}. However, an important metric that remains largely unexplored in the literature, due to its probabilistic nature and the inherent complexity of the associated optimization problem, is the probability of outage. Outage-constrained DFRC systems are studied in \cite{bazzi2023outage, xiong2024robust, soltani2023outage}. 
The models in \cite{bazzi2023outage, xiong2024robust} ignore multiple targets: 
\cite{bazzi2023outage} maximizes power in one direction, while \cite{soltani2023outage} introduces a target outage probability. 
In contrast, \cite{xiong2024robust} considers multiple targets by maximizing radar SINR, treating them similarly to communication users. 
We instead apply outage probability to multiple targets, accounting for their cross-correlation. 
While \cite{bazzi2023outage} uses the Bernstein inequality for probabilistic constraints, we employ the Gaussian error function and define two objective functions for optimization.

The majority of radar systems operate in a mono-static configuration, utilizing the same antenna for both transmission and reception. In such setups, pulsed signals are well-suited, allowing the radar to transmit during one time interval and receive the echo during another. This approach eliminates the need for additional isolation between the transmitter and receiver, as only one of them is active during each time period. The use of a pulsed system poses challenges for a communication-centric design because communication is halted during the radar's listening time. Therefore, the system does not operate in full duplex (FD). Utilizing continuous-wave (CW) to achieve FD signaling increases the risk of self-interference in a mono-static configuration. A bi-static system offers an advantage in this regard, as the receiver and transmitter are not collocated and employ distinct sets of antennas. But this in turn increases the cost of hardware. Consequently, the disparities in communication and radar signals impose constraints on the integration of these two systems, see, e.g., \cite{zhang2021overview}. Acknowledging the challenges mentioned above, a considerable research effort has been directed towards addressing these issues. 

An important factor in the design of DFRC is the trade-off between radar (or sensing) and communication performance. Focusing too much on the radar aspect can deteriorate communication performance, and vice versa. Therefore, a balance between the two should be achieved. 
A weighted optimization approach to achieve a flexible balance using low-complexity global algorithms to derive closed-form solutions for dual-function waveform designs is proposed in \cite{yang2023towards}.
Other papers that consider the tradeoff between communication and radar performance include \cite{liu2020range, ieee2023multiuser, ieee2021limited}. 

In this paper, we consider beamforming optimization in both radar-centric and communication-centric DFRC scenarios. While existing literature has addressed the problems of beamforming, outage-constrained communications, and radar waveform design independently, our work presents a unified framework that integrates these elements for DFRC systems and presents a joint formulation and its tractable optimization-based solution. Our main contributions are as follows:
 
\begin{itemize}
\item In the radar-centric scenario, we aim to match a desired radar beampattern while minimizing mean-squared cross-correlation among radar returns at multiple directions of interest (DOIs). Having multiple DOIs and accounting for cross-correlation lead to objectives and solutions differing from single-DOI works, where cross-correlation is irrelevant. To accommodate communication users, we impose outage probability constraints for each user. The optimization variables are the beamforming vectors for forward-channel users, with the transmit signal confined to their span. We convert probabilistic constraints to deterministic ones and apply a semi-definite relaxation (SDR). Since Gaussian randomization is often unsuitable or infeasible for this SDR, we employ a penalty-based approach.


\item For the communication-centric scenario, the main goal is to minimize the maximum probability of outage among users. 
To accommodate the radar system, we make sure that the radar performance metrics, beampatern matching error and cross-correlation between different DOIs, remains below a pre-specified threshold. This scenario is particularly relevant for passive radar, where reflections from the communication signal are used to monitor prespecified directions of interest. The resulting problem is non-convex, and we resort to a bisection method combined with a penalty term to handle the rank-1 constraint in the SDR formulation and the quasi-convex constraint.
\item In each case, our formulation of the semidefinite program relies on invoking a generalized version of the central limit (CLT) theorem with dependent variables to express outage probabilities as conic functions of the beamforming decision variable. We outline the conditions required for the validity of this framework and demonstrate through simulations that the CLT remains applicable even with a small number of antennas.

\end{itemize}
We also evaluate the performance of the proposed solutions through numerical simulations and show the validity of the proposed methods.

The remainder of the paper is organized as follows. Section \ref{SysMod} presents the system model and introduces the performance metrics relevant to communication and radar. In Sections \ref{radar-centric} and \ref{comm-centric}, we study the bemaforming problem in the radar-centric and the communication-centric scenarios, respectively. Section \ref{Results} presents our simulation results for the proposed methods. Finally, in Section \ref{conclusion} we will provide our conclusions and propose directions for future research.

\section{System Model}
\label{SysMod}

\begin{figure}
    \centering
    \includegraphics[width = 0.45\textwidth,trim=5.5cm 0cm 10cm 0cm, clip=true]{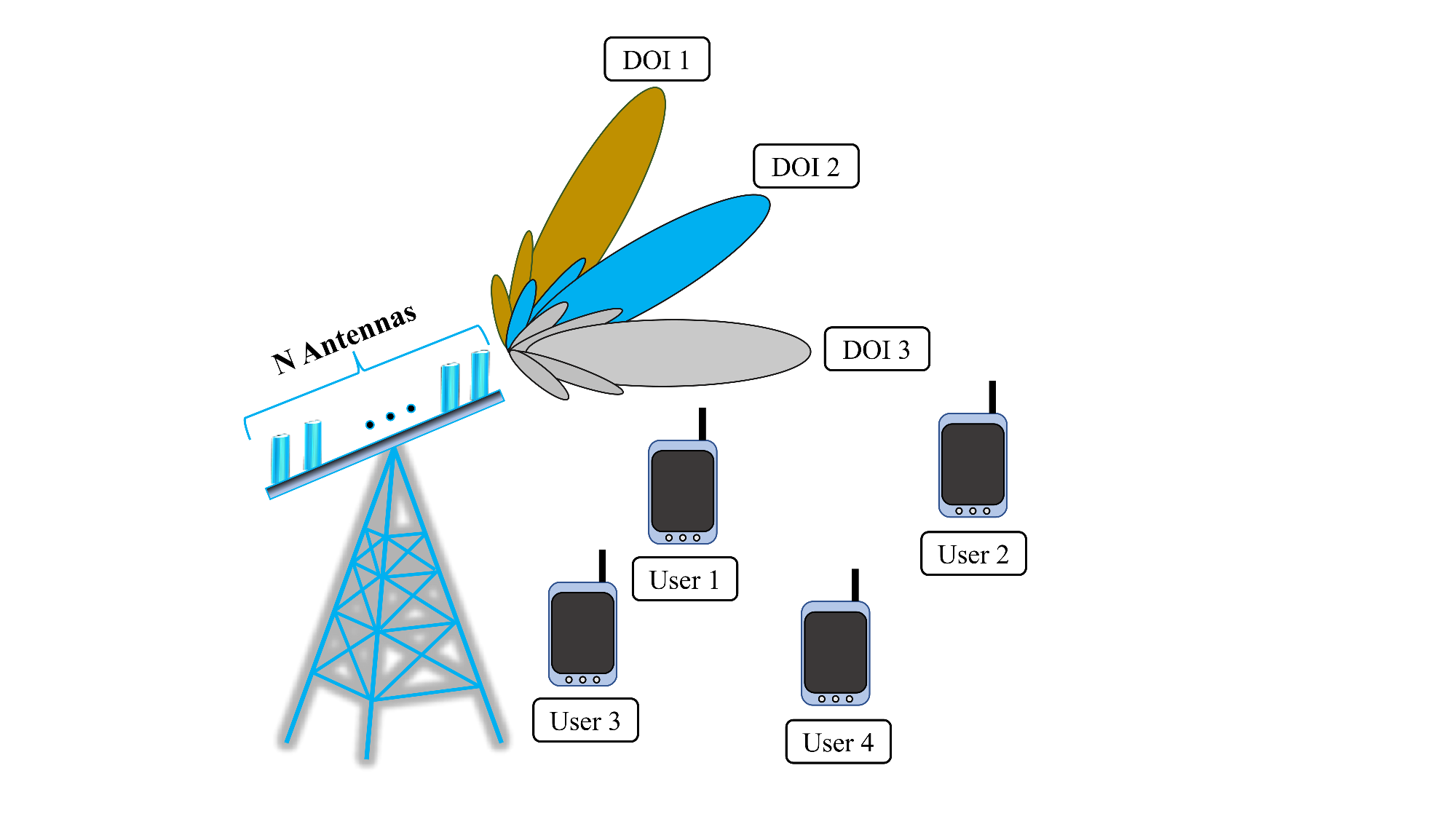} 
    \caption{A DFRC system with $K$ downlink communication users and $M$ radar directions of interest}
    \label{fig:fig1}
\end{figure}

We consider a collocated MIMO DFRC system equipped with a single transmitter with $N$ transmit antennas configured as a uniform linear array (ULA), $M$ azimuth directions of interest,  and $K$ single-antenna downlink users, as shown in Fig. \ref{fig:fig1}. 
The system employs the same antenna array for both transmission and reception in a time division (TD) manner. 
Complex zero-mean symbols 
$s_1, \dots, s_K$ are transmitted by the transmitter to the forward channel users with power $P_T$. Furthermore, it is assumed that the data symbols have unit energy and are uncorrelated, i.e., $\mathbb{E}[|s_k|^2] = 1$ and $\mathbb{E}[s_k s^*_j]=0, k\neq j$. Hence, the transmitted signal by the BS is 
\begin{equation}
    \label{eqn: transmit_sig}
    \mathbf{x} = \sum_{k=1}^K\mathbf{w}_k s_k,
\end{equation}
where $\mathbf{w}_k$ is the beamforming vector associated to User $k$. Given the dual purpose of the system, the same waveform is used for the radar functionality. The radar echo signals are recorded during the reception mode to estimate the parameters of interest.

Below, we will describe the communication and radar operations of the system in separate subsections. 

\subsection{Communication System}


Let $\mathbf{h}_k$ represent the zero-mean Gaussian channel vector between the BS and User $k$ with covariance matrix $\mathbf{C}_k = \mathbb{E}[\mathbf{h}_k\mathbf{h}_k^H]$. We assume that the channel covariance matrix is not known exactly and instead, an estimated version is available. The effects of imperfect channel state information (CSI) because of non-ideal estimation or quantized feedback in communication systems with beamforming have been studied in the literature \cite{HJ05,Sabharwal,EkJa,6172678,cui2018outage,EKHJ14}. 
 However, the effects of imperfect CSI on DFRC beamforming  seems to be less well-studied.

 There are different methods for estimation of the channel covariance matrix such as those proposed in \cite{Khalilsarai2019, Khalilsarai}, which is not the focus of this paper. We denote by $\hat{\mathbf{C}}_k\in \mathbb{C}^{N\times N}$  the estimate of the channel covariance matrix for user $k$.  Additionally, $\mathbf{E}_k \in \mathbb{C}^{N\times N}$ represents the uncertainty in the estimate of the covariance matrix, whose upper-triangular elements are distributed according to some distribution with mean $\mu_{k,ij}$ and variance $\sigma_{k,ij}^2$, and the lower-triangular elements are the conjugates of the upper-triangular elements. Therefore, the true covariance matrix satisfies $\mathbf{C}_k =  \hat{\mathbf{C}}_k + \mathbf{E}_k$. 

Upon transmitting the signal in Eq. \eqref{eqn: transmit_sig}, the received signal at User $k$ is
\begin{equation}
    \label{eqn: rec_comm}
    y_k = \sum_{j=1}^K \mathbf{h}_k^H\mathbf{w}_j s_j + n_k, \quad \forall k \in \{1, \dots, K\},
\end{equation}
where $n_k \sim \mathcal{CN}(0, \sigma^2)$ is the additive white complex Gaussian noise. The SINR at User $k$ can be written as
\begin{equation}
    \label{eqn: SINR}
    \textrm{SINR}_k = \frac{\mathbb{E}[|\mathbf{h}_k^H \mathbf{w}_k|^2]}{\sum_{j\neq k}\mathbb{E}[|\mathbf{h}_k^H \mathbf{w}_j|^2] + \mathbb{E}[|n_k|^2]}.
\end{equation}
Using $|v|^2 = v^*v$, Eq. \eqref{eqn: SINR} can be rewritten as
\begin{equation}
    \label{eqn: SINR2}
    \textrm{SINR}_k = \frac{\mathbf{w}_k^H \mathbb{E}[\mathbf{h}_k \mathbf{h}_k^H]\mathbf{w}_k}{\sum_{j\neq k}\mathbf{w}_j^H \mathbb{E}[\mathbf{h}_k \mathbf{h}_k^H]\mathbf{w}_j + \sigma^2}.
\end{equation}
Substituting the expression for the channel covariance matrix, the SINR will be
\begin{equation}
    \label{eqn: SINR3}
    \textrm{SINR}_k = \frac{\mathbf{w}_k^H (\hat{\mathbf{C}}_k + \mathbf{E}_k)\mathbf{w}_k}{\sum_{j\neq k}\mathbf{w}_j^H (\hat{\mathbf{C}}_k + \mathbf{E}_k)\mathbf{w}_j + \sigma^2}.
\end{equation}

Let us consider the event that the outage probability of communication User $k$ is not exceeding a prespecified threshold $p_k$. The probability of this event is bounded below by  
\begin{equation}
    \label{eqn: p_outage}
    \mathrm{Pr}[\mathrm{SINR}_k \geq \gamma_k] \geq 1-p_k, \quad \forall k \in \{1, \dots, K\},
\end{equation}
and guarantees the reliability of communication to the degree of interest. Outage probability is an effective performance measure for systems that incorporate uncertainty. Since the channel covariance matrices are accompanied by an error matrix, we adopt this metric to evaluate the performance of the communication component in both radar-centric and communication-centric tasks.

\subsection{Radar System}
With the assumption of narrow-band radar and line-of-sight (LoS) propagation, the baseband signal at angular direction $\theta$ can be expressed as
\begin{equation}
    \label{eqn: radar_signal}
    r(\theta,d) =\beta(d) \mathbf{a}^H(\theta) \mathbf{x},
\end{equation}
where $\mathbf{a}(\theta) \triangleq [1, e^{j\frac{2\pi}{\lambda}\Delta\sin \theta}, \dots, e^{j\frac{2\pi}{\lambda}(N-1)\Delta\sin \theta}]^T$ is the steering vector for the angular direction $\theta$, $\lambda$ is the wavelength, $d$ is the distance, $\beta(d)$ is the path-loss coefficient, and $\Delta$ is the spacing between the elements of the ULA. The power of beampattern at angular direction $\theta$ and distance $d$ is given by
\begin{align}
    P(\theta, d) &= \mathbb{E}[ \big|r(\theta, d)\big|^2] =  \mathbb{E}[\beta^2(d)\mathbf{a}^H(\theta) \mathbf{x} \mathbf{x}^H \mathbf{a}(\theta)] \nonumber \\ &= \sigma^2_d \mathbf{a}^H(\theta) \mathbf{R} \mathbf{a}(\theta),
\end{align}
where $\mathbf{R} = \mathbb{E}[ \mathbf{x} \mathbf{x}^H]$ is the covariance of transmit waveform and $\sigma^2_d$ is the path-loss power.  Substituting $\mathbf{x}$ from Eq. \eqref{eqn: transmit_sig}, we have $\mathbf{R}= \sum_{k=1}^K\mathbf{w}_k \mathbf{w}_k^H$. 

We wish to match the radar beampattern to a desired response at a prespecified set of angles, which constitute directions of interest for radar surveillance. At the same time, we wish to reduce the cross-correlation of the radar echos at these angles to control the sidelobe leakage from one angle to another.
The combination of these two criteria provides an objective function for the performance of a radar system with multiple directions of interest \cite{li2007mimo, stoica2007probing}. 

The first criterion, corresponding to the beampattern matching error, is defined as the mean square error (MSE) between the achieved beampattern and a given desired beampattern. We look at the radar beampattern as a function of the azimuth angle, and at a fixed range. Hence, without loss of generality, we set $\sigma^2_d$ equal to $1$. That is,
\begin{equation}
    \label{integral_MSEE}
    L_1(\mathbf{R}, \alpha) = \frac{1}{\pi} \int_{-\frac{\pi}{2}}^{\frac{\pi}{2}} \left[\alpha \phi(\theta)-\mathbf{a}^H(\theta) \mathbf{R} \mathbf{a}(\theta)\right]^2 \mathrm{d}\theta,
\end{equation}
where $\alpha > 0$ is a scaling factor, $\phi(\theta)$ is the desired beampattern in the angular direction $\theta$. This integral can be approximated by a summation when we regularly discretize the angular direction into $L$ points,  $\{\theta_l\}_{l=1}^L$:
\begin{equation}
    \label{eqn: loss1}
    L_1(\mathbf{R}, \alpha) = \frac{1}{L} \sum_{l=1}^L \left[\alpha \phi(\theta_l)-\mathbf{a}^H(\theta_l) \mathbf{R} \mathbf{a}(\theta_l)\right]^2.
\end{equation}

To define the second criterion, we first need to define the cross correlation between radar responses at different angular directions. The cross correlation at directions $\theta_i$ and $\theta_j$ is defined as
\begin{align}
    R_c(\theta_i, \theta_j;R) &= \mathbb{E}[r(\theta_i)r^*(\theta_j)]=\mathbb{E}[\mathbf{a}^H(\theta_i)\mathbf{x}\mathbf{x}^H \mathbf{a}(\theta_j)] \nonumber \\ &= \mathbf{a}^H(\theta_i) \mathbf{R} \mathbf{a}(\theta_j).
\end{align}
To see the importance of this measure, note that the echo signal received at radar will be
\begin{equation}
    \mathbf{y}_R = \sum_{j\in \mathcal{T}}\kappa_j \mathbf{a}(\theta_j)\mathbf{a}^H(\theta_j)\mathbf{x} + \mathbf{n}_R,
\end{equation}
where $\mathcal{T}$ shows the set of DOIs, $\kappa_j$ denotes the round trip loss which depends on parameters such as distance of the target and radar cross section (RCS), and $\mathbf{n}_R$ captures noise and clutter. 
For detection in the direction of interest $\theta_i$, the received signal at the radar will be multiplied by a filter $\mathbf{a}(\theta_i)$. That is
\begin{equation}
    z(\theta_i) = \mathbf{a}^H(\theta_i) \mathbf{y}_R = \sum_{j\in \mathcal{T}}\kappa_j  \mathbf{a}^H(\theta_i)  \mathbf{a}(\theta_j)\mathbf{a}^H(\theta_j)\mathbf{x} +  \mathbf{a}^H(\theta_i)\mathbf{n}_R.
\end{equation}
Magnitude of $z(\theta_i)$, $\lvert z^2(\theta_i)\rvert$, gives us the statistics for detection. Calculating $z(\theta_i)z^*(\theta_i)$ results in
\begin{equation}
   \lvert z^2(\theta_i)\rvert = \sum_{j, l \in \mathcal{T}} \kappa_j \kappa_l^* 
c_{ij}
c_{il}
\mathbf{a}^H(\theta_j) \mathbf{R} \mathbf{a}(\theta_l) + n_i,
\end{equation}
where $c_{ij} = \mathbf{a}^H(\theta_i)\mathbf{a}(\theta_j)$ and $n_i$ is the corresponding noise and clutter power. Therefore, we are interested in minimizing the cross correlation terms that appear as $\mathbf{a}^H(\theta_j) \mathbf{R} \mathbf{a}(\theta_l)$.
Based on this definition, the second criterion that captures cross-correlation among all pairs of angles of interest can be expressed as
\begin{equation}
    \label{eqn: loss2}
    L_2(\mathbf{R}) = \frac{2}{M^2 - M}\sum_{m=1}^{M-1}\sum_{n=m+1}^M \left|\mathbf{a}^H(\Tilde{\theta}_m) \mathbf{R} \mathbf{a}(\Tilde{\theta}_n)\right|^2,
\end{equation}
where $\{\Tilde{\theta}_m\}_{m=1}^M$ are the given directions of interest.
There are $\frac{M^2 - M}{2}$ distinct pairs of angles. Hence, the loss function is normalized by the number of pairs in the summation. 

The radar loss function is a weighted sum of $L_1(\mathbf{R}, \alpha)$ and $L_2(\mathbf{R})$. That is, 
\begin{equation}
    \label{loss}
    L(\mathbf{R}, \alpha) = L_1(\mathbf{R}, \alpha) + \delta L_2(\mathbf{R}).
\end{equation}
This loss function was proposed in \cite{stoica2007probing} for radar signal design only, without considering a dual radar and communication design problem, which is the goal of our paper. 


\section{Radar-centric Beamforming Problem}\label{radar-centric}

Combining the requirements of the communication and  radar systems, we formulate the following optimization problem:
\begin{mini!}<b>
{\substack{\alpha, \{\mathbf{w}_k\}_{k=1}^K}}{L(\mathbf{R}, \alpha)}
{\label{opt: 1}}{}
\addConstraint{
  \begin{aligned}
    &\mathrm{Pr}[\mathrm{SINR}_k \geq \gamma_k] \geq 1-p_k \\
    &\quad \forall k \in \{1, \dots, K\}
  \end{aligned}
}{}{\label{constraint13b}}
\addConstraint{[\mathbf{R}]_{n,n}=\frac{P_T}{N}, \quad \forall n \in \{1, \dots, N\}.}
\end{mini!}

The first constraint guarantees the probability of outage for forward channel communication users remains within the required limits. The second constraint ensures automatic satisfaction of the total power budget constraint with the same power for different antennas. Using equal power across antennas ensures that all amplifiers work near their optimal operating point to maximize efficiency and minimize distortion or nonlinear effects.

It is shown in \cite{stoica2007probing} that $L(\mathbf{R}, \alpha)$ can be written as a positive semi-definite quadratic function of $\mathbf{R}$ and $\alpha$, and semi-definite quadratic programming (SQP) can be used to minimize it without constraints.
However, the first constraint is not deterministic and requires modifications to make it tractable. 
In \cite{chalise2007robust}, it is assumed that the error matrix entries are independent and Gaussian to transform the first constraint to an equivalent deterministic one. In this work, we consider more general distributions for the entries of the error matrix, i.e., relax the Gaussian and independence assumptions.
Returning to Eq. \eqref{eqn: SINR3}, using $\mathrm{Tr}[\mathbf{AB}] = \mathrm{Tr}[\mathbf{BA}]$ and the fact that a scalar is equal to its trace, we can rewrite the SINR as
\begin{align}
\label{eqn: SINR4}
\textrm{SINR}_k &= \frac{\mathrm{Tr}[\mathbf{w}_k^H (\hat{\mathbf{C}}_k + \mathbf{E}_k)\mathbf{w}_k]}{\sum_{j\neq k}\mathrm{Tr}[\mathbf{w}_j^H (\hat{\mathbf{C}}_k + \mathbf{E}_k)\mathbf{w}_j] + \sigma^2} \nonumber \\ 
&= \frac{\mathrm{Tr}[\mathbf{w}_k\mathbf{w}_k^H (\hat{\mathbf{C}}_k + \mathbf{E}_k)]}{\sum_{j\neq k}\mathrm{Tr}[\mathbf{w}_j\mathbf{w}_j^H (\hat{\mathbf{C}}_k + \mathbf{E}_k)] + \sigma^2} \nonumber \\ 
&= \frac{\mathrm{Tr}[\mathbf{W}_k (\hat{\mathbf{C}}_k + \mathbf{E}_k)]}{\sum_{j\neq k}\mathrm{Tr}[\mathbf{W}_j (\hat{\mathbf{C}}_k + \mathbf{E}_k)] + \sigma^2},
\end{align}
where $\mathbf{W}_k = \mathbf{w}_k\mathbf{w}_k^H, \forall k \in \{1, \dots, K\}$. Substituting the SINR expression in the outage probability constraint gives
\begin{align}
\label{eqn: prob_trace}
&\mathrm{Pr}[\mathrm{Tr}[-\mathbf{B}_k\mathbf{E}_k] \leq \mathrm{Tr}[\mathbf{B}_k\mathbf{\hat{C}}_k]-\sigma^2] \geq 1- p_k,\\
&\forall k \in \{1, \dots, K\}, \nonumber
\end{align}
where $\mathbf{B}_k = \frac{1}{\gamma_k}\mathbf{W}_k - \sum_{j\neq k} \mathbf{W}_j$. Without loss of generality, we assume that the means of the elements of  the error matrix $\mathbf{E}_k$, denoted by $\mu_{k,ij}$, are zeroes (otherwise, we can define a mean matrix $\boldsymbol{M}_k$  and let $\hat{\mathbf{C}}_k^{'} = \hat{\mathbf{C}}_k + \boldsymbol{M}_k$). 
Our goal is to find a deterministic expression for this probabilistic constraint. To this end, we need to find the distribution of the random variable $\mathrm{Tr}[-\mathbf{B}_k\mathbf{E}_k]$. Note that this is not the only possible approach for solving the problem, and other transformations and approximations may be used, but finding the distribution as we propose results in exact bounds. We consider two different uncertainty models. In the first model, elements of $\mathbf{E}_k$ are assumed to be mutually independent. This uncertainty model for the covariance (with the additional assumption of Gaussianity of its elements) has been used in several prior works such as \cite{Alavi2018, Nasseri2013}. As the number of antennas increases, the number of added elements in the trace term increases as well.  Therefore, we can apply CLT and show that $\mathrm{Tr}[-\mathbf{B}_k\mathbf{E}_k]$ converges in distribution to a Gaussian as $N \rightarrow \infty$.   For this model, we assume that the upper-triangular elements of $\mathbf{E}_k$ are independent. Clearly, the lower triangular elements are not independent of the upper-triangular ones, as they are the conjugate of the upper-triangular elements. Applying the Lyapunov CLT theorem for independent random variables as outlined in Appendix \ref{App:A} gives us the desired result.\footnote{In our analysis, we fix the number of users $K$ and let $N \rightarrow \infty$. The analysis holds when $K << N$. Later, we show empirically that the result is also valid for small values of $N$.}
Next, we need to find the variance of this zero-mean random variable. 
For $i<j$, we can write $e_{k, ij} = y_{k, ij} + jz_{k, ij}$ and $e_{k, ji} = y_{k, ij} - jz_{k, ij}$ where $e_{k, ij}$ is the {\it ij}th element of $\mathbf{E}_k$ and $y_{k, ij}$ and $z_{k, ij}$ are independent zero-mean random variables with variances $\sigma_{k, ij}^2/2$. We are interested in the variance of $\mathrm{Tr}[-\mathbf{B}_k\mathbf{E}_k]$ which we can write as

\begin{align}
    \mathrm{Tr}[-\mathbf{B}_k\mathbf{E}_k] &= 
    \underbrace{-\sum_{j=1}^n \sum_{i=j}^n b_{k,ij}e_{k,ji}}_{\text{upper triangle}} 
    - 
    \underbrace{\sum_{j=2}^n \sum_{i=1}^{j-1} b_{k,ij}e_{k,ji}}_{\text{lower triangle}}\nonumber \\&= 
    -\sum_{i=1}^n b_{k,ii}e_{k,ii} - \sum_{j=1}^{n-1} \sum_{i=j+1}^{n} 2\mathrm{Re}\{b_{k,ij}\}y_{k,ji}\nonumber \\ &+ \sum_{j=1}^{n-1} \sum_{i=j+1}^{n}2\mathrm{Im}\{b_{k,ij}\} z_{k,ji}. 
\end{align}
Since this is a zero-mean random variable, the variance $\sigma^2$ can be calculated as
\begin{align}
    \sigma^2 &= \sum_{i=1}^n b_{k,ii}^2\sigma_{k,ii}^2 + \sum_{j=1}^{n-1} \sum_{i=j+1}^{n} 4\mathrm{Re}^2\{b_{k,ij}\}\frac{\sigma_{k,ji}^2}{2} \\ &+ \sum_{j=1}^{n-1} \sum_{i=j+1}^{n}4\mathrm{Im}^2\{b_{k, ij}\}\frac{\sigma_{k,ji}^2}{2}\nonumber \\
     &=\sum_{i=1}^n b_{k,ii}^2\sigma_{k,ii}^2 + \sum_{j=1}^{n-1} \sum_{i=j+1}^{n} 2\lvert b_{k,ij}\rvert^2 \sigma_{k,ji}^2 \nonumber \\ &= \sum_{i=1}^n \sum_{j=1}^n |b_{k,ij}|^2\sigma_{k,ij}^2 = \lVert \mathbf{B}_k \odot\boldsymbol{\Sigma}_k\rVert_F^2, \nonumber
\end{align}
where $\mathbf{\Sigma}_k$ contains the variances of the elements of $\mathbf{E}_k$, $\lVert  \rVert_F$ is the Frobenius norm, and we used the fact that $\lvert b_{k,ij}| = \lvert b_{k,ji}\rvert$. In the second scenario, we consider a more general uncertainty model where elements of $\mathbf{E}_k$ are dependent, but their correlation vanishes as a function of covariance lag. In Appendix \ref{App: B}, we show that by employing a generalized version of the central limit theorem, we can still establish that $\mathrm{Tr}[-\mathbf{B}_k\mathbf{E}_k]$ tends to a Gaussian distribution. We note that there are several different versions of the generalized central limit theorem that we can use to establish this result, each with slightly different assumptions. The interested reader can refer to \cite{Dudley_1999} for a thorough study of different generalized central limit theorems. In Appendix \ref{App: B}, we only present one such result. In a nutshell, if the dependence between the random variables decays as the difference between their indices increases, then the CLT holds. The elements of the error matrix in our study model the errors in estimation of the covariance between the channels corresponding to two antennas. We know that as the antennas become further separated, the correlation between the corresponding channels decreases. Therefore, we can assume that the dependence between the elements of the error matrix that correspond to largely separated antennas (large difference in indices) is small. The specific indexing of the random variables that we use (sorting them diagonal by diagonal) ensures that the spatial distance between the corresponding antenna elements increases as the corresponding index increases. This is because, within the Hermitian matrix, the diagonal corresponds to the smallest spatial distance ($ i = j $), and as we move to elements in the upper triangle grouped by increasing diagonal offsets ($ j - i $), the spatial separation between the indices $i$ and $j$ grows. This ordering reflects a systematic progression from the closest to the farthest antenna elements. It only remains to find the  variance of this zero-mean random variable. 

Let $\boldsymbol{\Gamma} = \mathbb{E}[\mathrm{vec}(\mathbf{E}_k )\mathrm{vec}(\mathbf{E}_k)^H]$ where $\mathrm{vec}(.)$ stacks the columns of a matrix to build a vector. Then, $\mathrm{var}[-\mathrm{Tr}[\mathbf{B}_k\mathbf{E}_k]]=\mathbf{b}_k^H \boldsymbol{\Gamma_k}\mathbf{b}_k$ where $\mathbf{b}_k = \mathrm{vec}(\mathbf{B}_k)$. Since $\boldsymbol{\Gamma}_k$ is positive semidefinite, we can write $\boldsymbol{\Gamma}_k = \Tilde{\boldsymbol{\Gamma}}_k \Tilde{\boldsymbol{\Gamma}}_k^H$. Therefore, $\mathrm{var}[-\mathrm{Tr}[\mathbf{B}_k\mathbf{E}_k]]=\lVert \Tilde{\boldsymbol{\Gamma}}^H \mathbf{b}_k \rVert^2$.


\subsection{Optimization Formulation and Solution}


Using the discussed Gaussianity results, Eq. \eqref{eqn: prob_trace} can be written in terms of the error function $\mathrm{erf}(x) = \frac{2}{\sqrt{\pi}}\int_0^x e^{-t^2} \mathrm{d}t$ as\footnote{The results for the independent case can be derived in a similar way.}
\begin{align}
&\frac{1}{2}\left[1 + \mathrm{erf}\left(\frac{\mathrm{Tr}[\mathbf{B}_k\mathbf{\hat{C}}_k]-\sigma^2}{\sqrt{2}\lVert\Tilde{\boldsymbol{\Gamma}}^H \mathbf{b}_k\rVert}\right)\right]\geq 1-p_k , \\&\forall k \in \{1, \dots, K\}. \nonumber
\end{align}
Equivalently, 
\begin{align}
    \lVert \Tilde{\boldsymbol{\Gamma}}^H \mathbf{b}_k \rVert \leq \varepsilon_k \left(\mathrm{Tr}[\mathbf{B}_k\mathbf{\hat{C}}_k]-\sigma^2\right),
\end{align}
where $\varepsilon_k = \frac{1}{\sqrt{2}\mathrm{erf}^{-1}(1-2p_k)}$. This is a second order cone (SOC). Finally, to be used in a convex optimization framework, the constraint is written as linear matrix inequalities (LMIs):
\begin{align}
    \mathbf{D}_k = 
    \begin{bmatrix}
        \varepsilon_k \left(\mathrm{Tr}[\mathbf{B}_k\mathbf{\hat{C}}_k]-\sigma^2\right) \mathbf{I}_{N^2} & \Tilde{\boldsymbol{\Gamma}}^H \mathbf{b}_k\\ \mathbf{b}_k^H\Tilde{\boldsymbol{\Gamma}}  & \varepsilon_k \left(\mathrm{Tr}[\mathbf{B}_k\mathbf{\hat{C}}_k]-\sigma^2\right) 
    \end{bmatrix}\succeq \mathbf{0}.
\end{align}
Replacing the probabilistic constraint with the equivalent deterministic constraint, the optimization problem \eqref{opt: 1} can be rewritten as
\begin{mini!}<b>
{\substack{\alpha, 
\{\mathbf{W}_k\}_{k=1}^K}}{L\left(\sum_{k=1}^K \mathbf{W}_k, \alpha\right)}
{\label{opt:Nonrank1}}{}
\addConstraint{\mathbf{D}_k \succeq 0, \quad \forall k \in \{1, \dots, K\}}{}{\label{const:outage}}
\addConstraint{\left[\sum_{k=1}^K \mathbf{W}_k\right]_{n,n}=\frac{P_T}{N}, \quad \forall n \in \{1, \dots, N\}{\label{eq:diag-power}}}
\addConstraint{\mathbf{W}_k \succeq 0, \quad \forall k \in \{1, \dots, K\} {\label{PSD}}}
\addConstraint{\mathrm{Rank}[\mathbf{W}_k] = 1}{},{\label{rank}}
\end{mini!}
where the set of constraints in  \eqref{PSD} and \eqref{rank} is added to guarantee  $\mathbf{W}_k = \mathbf{w}_k \mathbf{w}_k^H$. This constraint makes the problem non-convex. 

A well-known method to solve this problem is to relax it by dropping the rank-1 constraint. The resulting convex optimization problem can be solved by CVX package \cite{cvx}. If the solution matrices are rank-1, they are the solutions to the original problem as well. If the solutions are not rank-1, randomization techniques can be employed to generate rank-1 solutions. The beamforming vectors, $\mathbf{w}_k$, will be determined through eigen-value decomposition of matrices $\mathbf{W}_k$. However, the equal per-antenna power constraint complicates the use of this method. We introduce a heuristic randomization method in the simulation results section for this problem, but it is only applicable to cases with mild constraints and channels. In complex settings with a large number of users, it usually does not provide decent candidates. 
Instead, we use the penalty method to enforce the rank-1 constraint \cite{wang2022noma}. A constraint equivalent to the rank-1 constraint is
\begin{equation}
    \lVert \mathbf{W}_K \rVert_* - \lVert \mathbf{W}_K \rVert_2 = 0,
\end{equation}
where $\lVert .\rVert_*$ denotes the nuclear norm (sum of the singular values of the matrix) and $\lVert .\rVert_2$ denotes the spectral norm (the largest singular value of the matrix). This constraint is non-convex too, but using the first order Taylor expansion for the spectral norm, we get a penalty term $P(\mathbf{W}_1, \dots, \mathbf{W}_k)$ that resolves the problem. The first-order Taylor expansion of the spectral norm is
\begin{align}
    \label{Taylor}
    -\lVert \mathbf{W}_k \rVert_2 \simeq f(\mathbf{W}_k) = & -\lVert \mathbf{W}_k^j \rVert_2 \\
    \nonumber
    & - \mathrm{Tr}[\mathbf{v}_{\mathrm{max},k}^j (\mathbf{v}_{\mathrm{max},k}^j)^H(\mathbf{W}_k - \mathbf{W}_k^j)],
\end{align}
where $\mathbf{W}_k^j$ is the point around which the Taylor expansion is calculated and $\mathbf{v}_{\mathrm{max},k}^j$ is the eigenvector corresponding to the largest
eigenvalue of $\mathbf{W}_k^j$. Small values of $ \lVert \mathbf{W}_K \rVert_* - \lVert \mathbf{W}_K \rVert_2$  indicate that the eigenvalues, except for the largest one, are very small. Therefore, ensuring a rank-1 solution is equivalent to ensuring that this difference is as close to zero as possible. We use the convex approximation and define a penalty term
\begin{equation}
    \label{penalty}
    P(\mathbf{W}_1, \dots, \mathbf{W}_k) = \sum_{k=1}^K  \lVert \mathbf{W}_K \rVert_* + f(\mathbf{W}_k).
\end{equation}
Since our optimization goal is to minimize the loss function defined in \eqref{loss}, and the rank-1 condition requires minimizing \eqref{penalty}, we can define a weighted objective function for optimization that takes both into account. The final optimization problem will be 
\begin{mini!}<b>
{\substack{\alpha, 
\{\mathbf{W}_k\}_{k=1}^K}}{L\left(\sum_{k=1}^K \mathbf{W}_k, \alpha\right) + \frac{1}{\zeta_j}P(\mathbf{W}_1, \dots, \mathbf{W}_k)}
{\label{Opt:final}}{}
\addConstraint{\mathbf{D}_k \succeq 0, \quad \forall k \in \{1, \dots, K\}}
\addConstraint{\left[\sum_{k=1}^K \mathbf{W}_k\right]_{n,n}=\frac{P_T}{N}, \quad \forall n \in \{1, \dots, N\}}
\addConstraint{\mathbf{W}_k \succeq 0, \quad \forall k \in \{1, \dots, K\}}.
\end{mini!}
The parameter \( \zeta_j \) controls the effect of the penalty in the objective function. Very large values of \( \zeta_j \) mean that the penalty is not important and mainly the loss is minimized. Conversely, small values of \( \zeta_j \) focus on the penalty and ensuring the rank-1 constraint, which can result in poor loss values.
We start with \( \zeta_0 = \infty \), which is equivalent to not incorporating the penalty term at all. This will give us the initial points, \( \mathbf{W}_k^1 \), for Taylor expansion. After that, we start with a large value of \( \zeta \) at the beginning and gradually decrease it by \( \zeta_j = \mu \zeta_{j-1} \) where \( 0 < \mu < 1 \) until we get rank-1 solutions.

\begin{algorithm}
\caption{Radar-Centric Beamforming Algorithm}\label{alg:a1}
\KwIn{Desired beampattern $\phi(\theta_l)$, SINR threshold $\gamma_k$, outage probability $p_k$, power budget $P_T$, number of transmit antennas $N$, CSI error covariance $\boldsymbol{\Gamma_k}$, $\zeta_1$ and $\mu$ for the penalty term, communication channel covariance matrix estimates $\hat{\mathbf{C}}_k$, and stopping criterion $\Delta$.}
\KwOut{Beamformers $\mathbf{w}_1, \dots, \mathbf{w}_K$}

\SetKwFunction{SolveOpt}{SolveOptimization}

\SetKwBlock{Initialization}{Initialization:}{}
\Initialization{
    Set $\zeta_0 = \infty$\;
    \SolveOpt{\eqref{Opt:final}}\;
    Store the solutions in $\mathbf{W}_k^1$\;
    Set the counter $j = 1$\;
}

\SetKwBlock{MainLoop}{Main Loop:}{}
\MainLoop{
    \While{$\lambda_2(\mathbf{W}_k) \geq \Delta$ for at least one $k$}{
        \SolveOpt{\eqref{Opt:final}}\;
        Store the solutions in $\mathbf{W}_k^{j+1}$\;
        $j = j + 1$\;
    }
}

Perform eigenvalue decomposition on $\mathbf{W}_k$ and set $\mathbf{w}_k = \sqrt{\lambda_1}\mathbf{v}_1$\;

\end{algorithm}

\section{Communication-centric Beamforming Problem}
\label{comm-centric}
In the previous section, we considered the radar-centric problem, in which the main goal was achieving the best radar performance while ensuring that communication users receive a minimum quality of service. In this section, we consider a dual problem. We aim to minimize the maximum probability of outage experienced by any user while ensuring that the radar performance remains acceptable. For the radar performance, we constrain the beampattern MSE defined in \eqref{eqn: loss1} and the cross correlation defined in \eqref{eqn: loss2} to corresponding threshold values. This scenario is particularly relevant for passive radar (see, e.g., \cite{chalise2017passive,hack2014passive} and the references therein), where reflections from the communication signal are used to monitor prespecified directions of interest.

By using the definition of probability of outage in \eqref{eqn: p_outage}, the desired optimization problem will be
\begin{mini!}<b>
    {\substack{\alpha, 
    \{\mathbf{W}_k\}_{k=1}^K}}{\max_k \mathrm{Pr}[\mathrm{SINR}_k \leq \gamma_k]}
    {\label{min_max1}}{}
    \addConstraint{L_1\left(\sum_{k=1}^K \mathbf{W}_k, \alpha\right)  \leq c_1{\label{L1_const}}}
    \addConstraint{L_2\left(\sum_{k=1}^K \mathbf{W}_k\right)  \leq c_2{\label{L2_const}}}
    \addConstraint{\left[\sum_{k=1}^K \mathbf{W}_k\right]_{n,n}=\frac{P_T}{N}, \quad \forall n \in \{1, \dots, N\}}
    {}{}
    \addConstraint{\mathbf{W}_k \succeq 0, \quad \forall k \in \{1, \dots, K\}}
\addConstraint{\mathrm{Rank}[\mathbf{W}_k] = 1}.
\end{mini!}

By introducing the auxiliary variable $t$, \eqref{min_max1} can be rewritten as

\begin{mini!}<b>
    {\substack{\alpha, t, 
    \{\mathbf{W}_k\}_{k=1}^K}}{t}
    {}{\label{min_max12}}
    \addConstraint{\mathrm{Pr}[\mathrm{SINR}_k \leq \gamma_k] \leq t}{}{\label{constraint25b}}
    \addConstraint{L_1\left(\sum_{k=1}^K \mathbf{W}_k, \alpha\right)  \leq c_1}
    \addConstraint{L_2\left(\sum_{k=1}^K \mathbf{W}_k\right)  \leq c_2}
    \addConstraint{\left[\sum_{k=1}^K \mathbf{W}_k\right]_{n,n}=\frac{P_T}{N}, \quad \forall n \in \{1, \dots, N\}}
    {}{}
    \addConstraint{\mathbf{W}_k \succeq 0, \quad \forall k \in \{1, \dots, K\}}
\addConstraint{\mathrm{Rank}[\mathbf{W}_k] = 1}{}.{\label{constrain25g}}
\end{mini!}
All constraints are convex except for \eqref{constraint25b} and \eqref{constrain25g}. While \eqref{constraint25b} is very similar to \eqref{constraint13b}, we should note that the right-hand side of \eqref{constraint13b} is a constant, while the right-hand side in \eqref{constraint25b} is an optimization variable. If we fix the variable $t$, then as we showed previously, we have a convex constraint. This type of constraint is called quasi-convex. Disregarding \eqref{constrain25g} for now, the idea to solve the problem is to use the bisection method. Since $t$ shows a probability, it is limited between 0 and 1. Therefore, we know the lower bound and the upper bound for $t$. We start by setting 
\begin{equation}
    \label{midpoint}
    t=\frac{\mathrm{upper + lower}}{2}
\end{equation}
and solving a feasibility problem, i.e.,  check the feasibility of constraints given the current value of $t$. If the problem is feasible, we set $\mathrm{upper} = t$ and recalculate $t$ using \eqref{midpoint}. Otherwise, we set $\mathrm{lower} = t$ and recalculate $t$ using \eqref{midpoint}. The algorithm stops when the difference between $\mathrm{lower}$ and $\mathrm{upper}$ is less than a threshold. That is $\mathrm{upper} - \mathrm{lower} < \epsilon$. This resolves the issue with constraint \eqref{constraint25b}. To rectify the problem with rank-1 constraints, we use the penalty method introduced in Section \ref{radar-centric}. To find an initial solution (which is not necessarily rank-1), we solve the feasibility problem. After that, we solve the following optimization problem
\begin{mini!}<b>
    {\substack{\alpha, 
    \{\mathbf{W}_k\}_{k=1}^K}}{P(\mathbf{W}_1, \dots, \mathbf{W}_k)}
    {\label{min_max3}}{}
    \addConstraint{\mathrm{Pr}[\mathrm{SINR}_k \leq \gamma_k] \leq t}{}{\label{constraint27b}}
    \addConstraint{L_1\left(\sum_{k=1}^K \mathbf{W}_k, \alpha\right)  \leq c_1}
    \addConstraint{L_2\left(\sum_{k=1}^K \mathbf{W}_k\right)  \leq c_2}
    \addConstraint{\left[\sum_{k=1}^K \mathbf{W}_k\right]_{n,n}=\frac{P_T}{N}, \quad \forall n \in \{1, \dots, N\}}
    {}{}
    \addConstraint{\mathbf{W}_k \succeq 0, \quad \forall k \in \{1, \dots, K\}}
\end{mini!}
where $t$ will be selected based on the bisection method. If the optimization problem becomes infeasible, we set $\mathrm{lower} = t$. If the problem is feasible and $\mathbf{W}_k$ are rank-1, we set $\mathrm{upper}=t$. If the problem is feasible, but $\mathbf{W}_k$ are not rank-1, we set $\mathrm{lower}=t$. Notice that the solver first makes sure that the constraints are satisfied and then minimizes the penalty. Therefore, it is possible that the constraints are satisfied, but there are no rank-1 solutions, and hence the value of penalty will be significant. In case that there exist rank-1 solutions, the solver finds it by minimizing the penalty term.

\begin{algorithm}
\caption{Communication-Centric Beamforming Algorithm}\label{alg:a2}
\KwIn{Desired beampattern $\phi(\theta_l)$, SINR threshold $\gamma_k$, power budget $P_T$, number of transmit antennas $N$, CSI error covariance $\boldsymbol{\Gamma_k}$, threshold on MSE error $c_1$ and threshold on cross-correlation error $c_2$, communication channel covariance matrix estimates $\hat{\mathbf{C}}_k$, eigenvalue stopping criterion $\Delta$ and bisection stopping criterion $\epsilon$.}
\KwOut{Beamformers $\mathbf{w}_1, \dots, \mathbf{w}_K$}

\SetKwFunction{SolveOpt}{SolveOptimization}

\SetKwBlock{Initialization}{Initialization:}{}
\Initialization{
    Set $\mathrm{upper} = 0.5$\;
    Set $\mathrm{lower} = 0$\;
    \While{$\mathrm{upper - lower} > \epsilon$}{
    $t=\frac{\mathrm{upper + lower}}{2}$\;
    Check the feasibility problem \eqref{min_max3} by dropping the penalty term\;
    \If{\eqref{min_max3} is feasible}{
        $\mathrm{upper} = t$\;
    }
    \Else{
        $\mathrm{lower} = t$\;
    }
    
    }
    Store the solutions in $\mathbf{W}_k^1$\;
    Set the counter $j = 1$\;
}

\SetKwBlock{MainLoop}{Main Loop:}{}
\MainLoop{
    Set $\mathrm{upper} = 0.5$\;
    Set $\mathrm{lower} = 0$\;
    \While{$\mathrm{upper - lower} > \epsilon$}{
    $t=\frac{\mathrm{upper + lower}}{2}$\;
    \SolveOpt{\eqref{min_max3}}\;
    \If{\eqref{min_max3} is feasible and $\lambda_2(\mathbf{W}_k) < \Delta$ for all $k$}{
        $\mathrm{upper} = t$\;
    }
    \Else{
        $\mathrm{lower} = t$\;
    }
    
    }
    Store the solutions in $\mathbf{W}_k^{j+1}$\;
}

Perform eigenvalue decomposition on $\mathbf{W}_k$ and set $\mathbf{w}_k = \sqrt{\lambda_1}\mathbf{v}_1$\;

\end{algorithm}

\section{Simulation Results}
\label{Results}
\begin{figure}[h]
    \centering
    \includegraphics[width=0.45\textwidth]{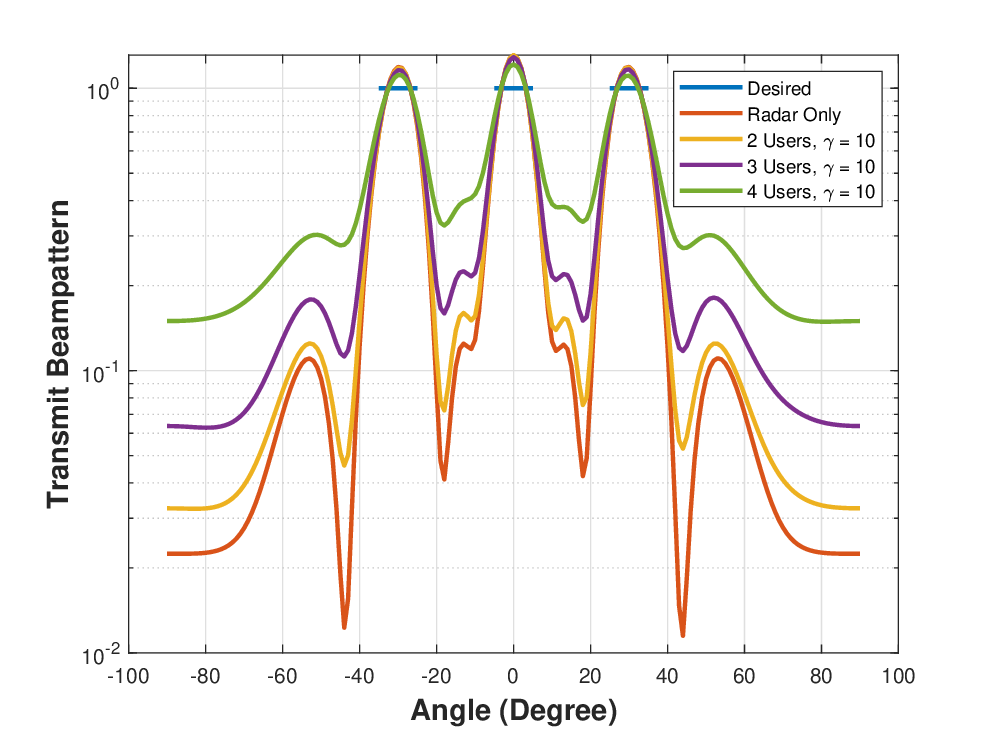}
    \caption{The ideal beampattern, the beampattern of radar-only system, and the beampattern of the proposed DFRC}
         
    \label{fig:Fig2}
\end{figure}

We assume a standard Rayleigh fading channel model for communication users. The carrier frequency is 5GHz and the ULA spacing is equal to $\lambda/2$. The transmitter's power budget and the thermal noise power are $P_T = 30 \mathrm{dBm}$ and $\sigma^2 = 10\mathrm{dBm}$, respectively. All figures are averaged over 100 Monte Carlo realizations of communication channels and the covariance estimates are calculated as $\mathbf{h}_k\mathbf{h}_k^H$. 
Figures that require different realizations of the error covariance matrix present averages over 1000 realizations.
Fig. \ref{fig:Fig2} shows the performance  for a transmitter equipped with $N=10$ antennas. In this setting, we assume the elements of the error matrix are independent Gaussian random variables and their variances are assumed to be the same and equal to $\sigma_e^2 = 0.005$. The probability of outage for all users is the same and equal to $p=0.1$. We considered 3 users at angles $-30$, $0$, and $30$ degrees, while we changed the number of communication users. Even though the outage threshold has been set to the large value of $\gamma_{th} = 10\mathrm{dB}$, we see a good match between the beampattern of the joint system and the radar-only system.  Adding more antennas to the system increases the degrees of freedom and reduces the mismatch between the beampatterns of the radar-only system and those of the DFRC. 

\begin{figure}[ht]
    \centering
    \begin{subfigure}[b]{0.45\textwidth}
        \includegraphics[width=\textwidth]{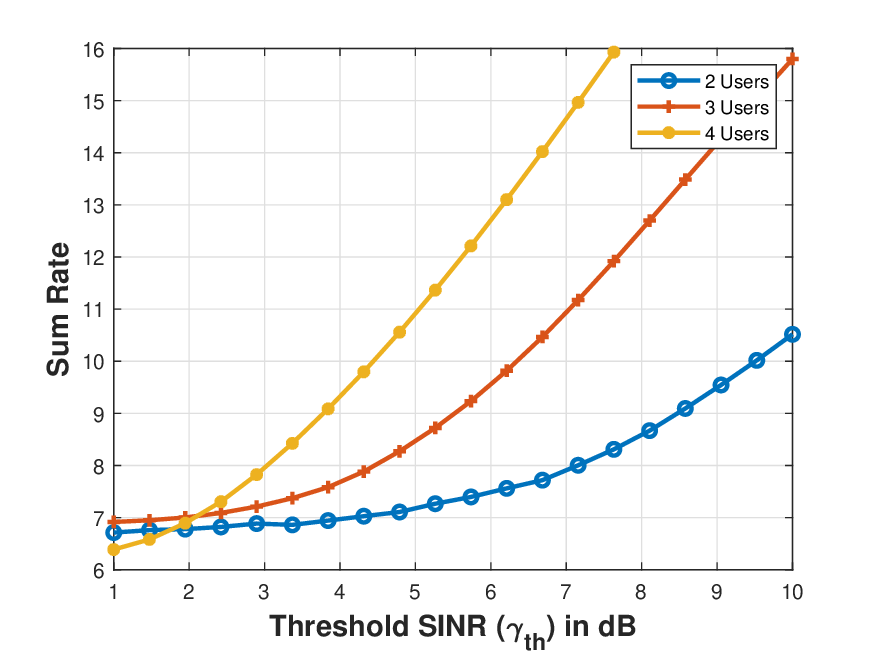}
        \caption{Sum rate (bps/Hz) versus different SINR thresholds}
        \label{fig:3a}
    \end{subfigure}
    \begin{subfigure}[b]{0.45\textwidth}
        \includegraphics[width=\textwidth]{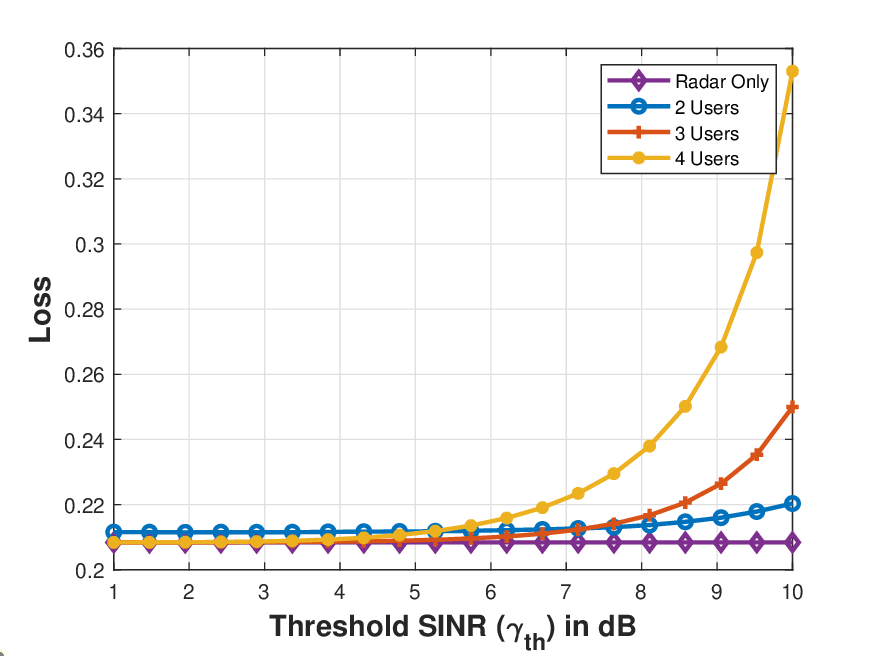}
        \caption{Combined loss versus different SINR thresholds, same setup as Fig.~\ref{fig:3a}}
        \label{fig: 3b}
    \end{subfigure}
    \caption{Performance of DFRC for fixed probability of outage ($p=0.1$) with different number of users}
    \label{fig:Fig3}
\end{figure}

Fig. \ref{fig:3a} illustrates the radar-centric scenario's sum rate with the outage probability of 0.1 for all users. Since the communication only occurs in the transmit interval of the radar, we present values scaled by the ratio of transmit plus listening time to transmit time to allow various designs of transmit and listening times. Fig. \ref{fig: 3b} shows the corresponding loss (combination of MSE and correlation loss) between the achieved beampattern and the ideal beampattern.
As observed from Fig.~\ref{fig:Fig3}, for a small number of users, the rise in the sum rate with an increase in the threshold SINR does not result in a substantial increase in the radar performance loss. This is potentially because as SINR threshold is increased  only slight adjustments to the beamforming vectors are needed. In systems with a large number of users, the system is constrained by interference, and achieving a higher SINR necessitates substantial changes in the beamforming vectors, consequently impacting radar performance. At low SINR thresholds, increasing the number of users introduces more interference and tighter beamforming constraints, which limits the system’s ability to maximize the sum rate. However, at higher SINR thresholds, the system can exploit multi-user diversity more effectively, allowing the sum rate to scale with the number of users.

\begin{figure}[ht]
    \centering
    \begin{subfigure}[b]{0.45\textwidth}
        \includegraphics[width=\textwidth]{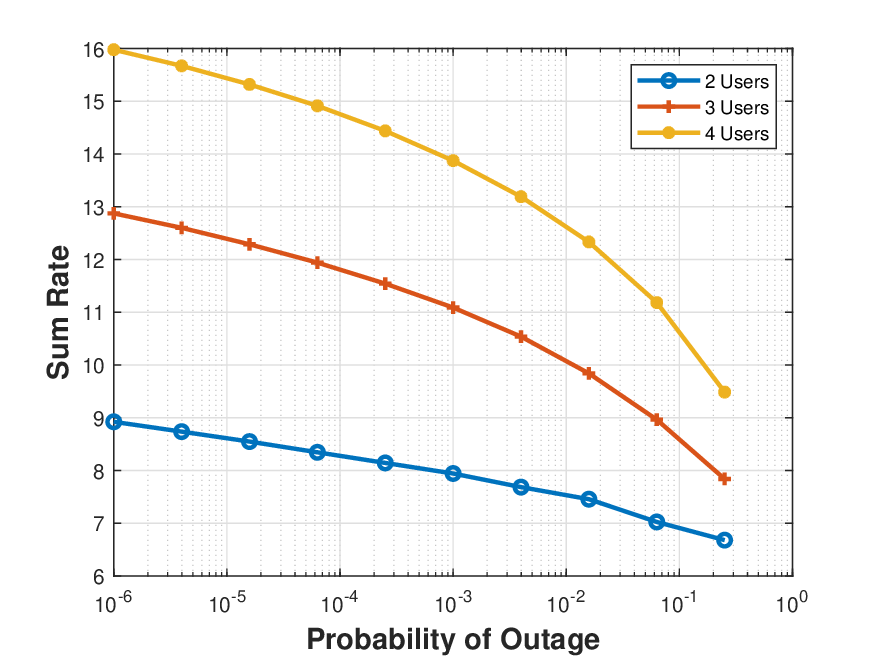}
        \caption{Sum rate (bps/Hz) versus probability of outage}
        \label{fig:4a}
    \end{subfigure}
    \begin{subfigure}[b]{0.45\textwidth}
        \includegraphics[width=\textwidth]{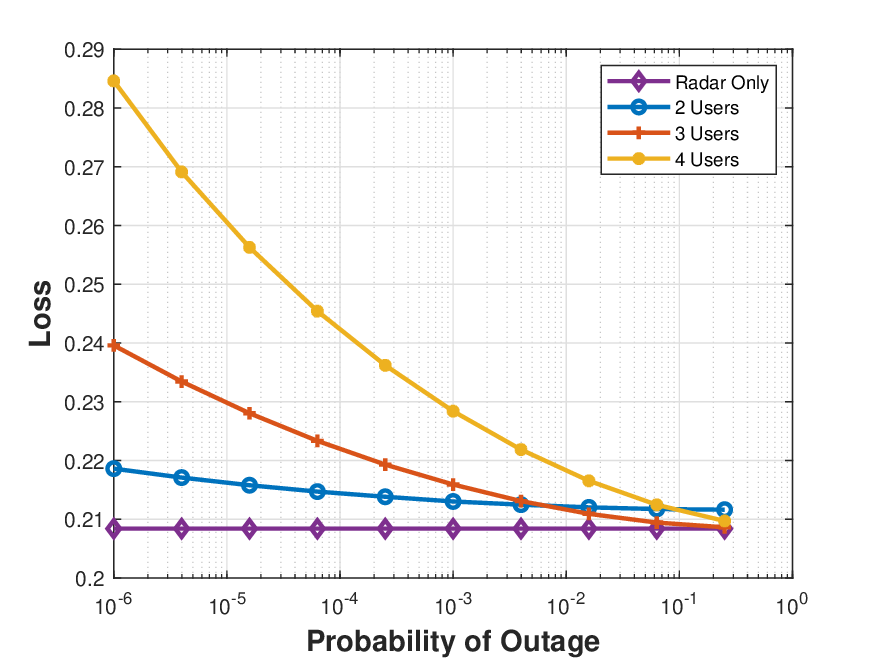}
        \caption{Combined loss versus probability of outage, same setup as Fig.~\ref{fig:4a}}
        \label{fig: 4b}
    \end{subfigure}
    \caption{Performance of DFRC for a fixed outage SINR threshold ($\gamma_{th}=5\mathrm{dB}$) with different number of users}
    \label{fig:Fig4}
\end{figure}

Fig. \ref{fig:Fig4} depicts the impact of the outage probability on DFRC performance. In this figure, the SINR threshold is set to $\gamma = 5\mathrm{dB}$ for all users and the system includes 3 radar directions of interest with different number of communication users. To study the tradeoff between the communication and radar performance, we have repeated the simulations for different design values of outage probabilities. With an increase in the acceptable outage probability, there is a decrease in the sum rate and in the mismatch between the desired and achieved beampatterns. This is intuitive, as a larger acceptable outage probability represents less importance placed on communication performance, resulting in improved radar performance.

\begin{figure}[ht]
    \centering
    \begin{subfigure}[b]{0.45\textwidth}
        \includegraphics[width=\textwidth]{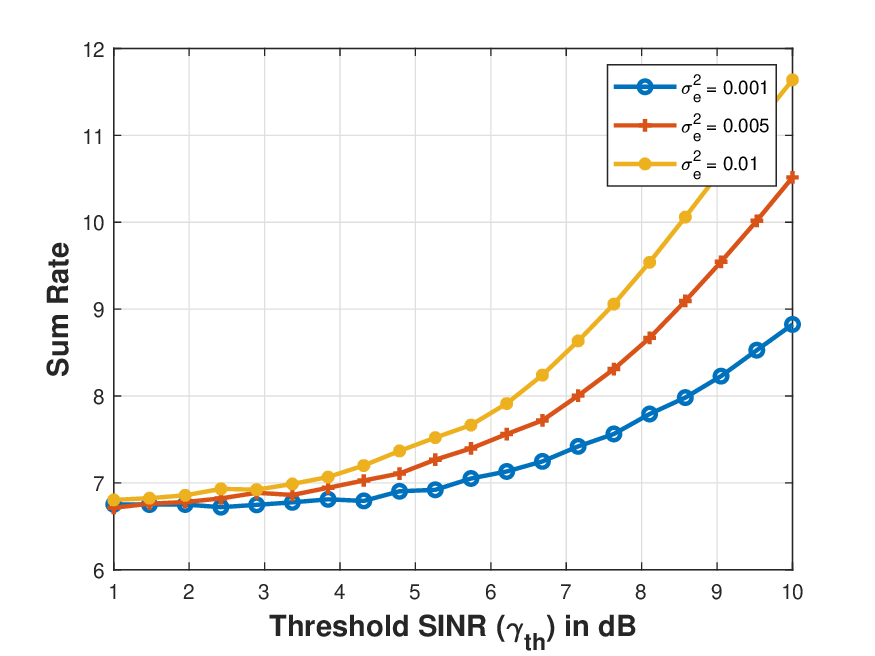}
        \caption{Sum rate (bps/Hz) versus different SINR thresholds}
        \label{fig:6a}
    \end{subfigure}
    \begin{subfigure}[b]{0.45\textwidth}
        \includegraphics[width=\textwidth]{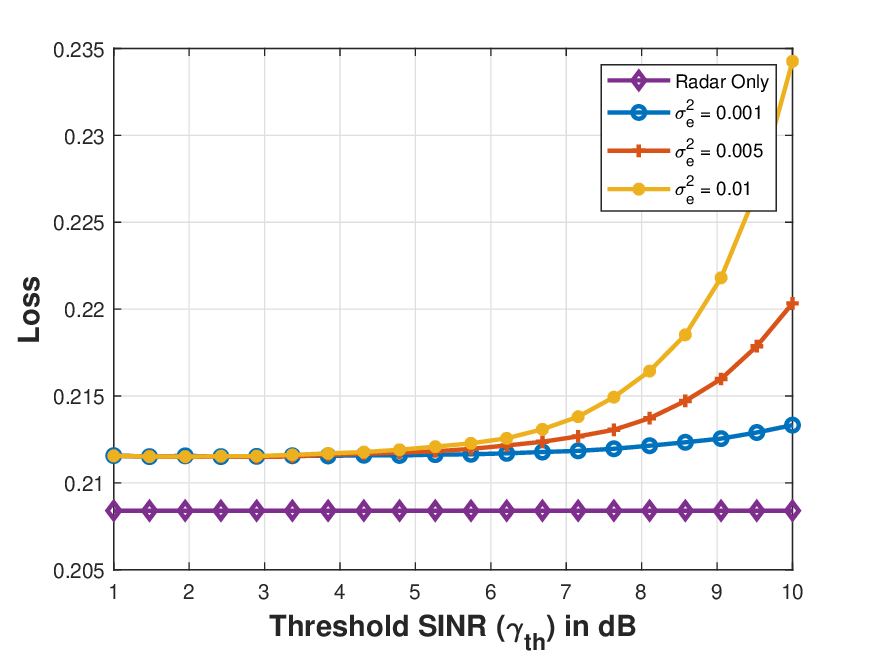}
        \caption{Loss for different values of SINR threshold, same setup as Fig.~\ref{fig:6a}}
        \label{fig: 6b}
    \end{subfigure}
    \caption{Performance of DFRC under different CSI uncertainty levels}
    \label{fig:Fig6}
\end{figure}

To study the effects of imperfect CSI, we have conducted the experiment with three distinct levels of uncertainty in channel estimation. Fig. \ref{fig:Fig6} illustrates that increased uncertainty in estimation leads to a larger loss (beampattern MSE and cross correlation) to fulfill a specific outage probability. In this setting, we assumed 2 users and 3 radar directions of interest. The probability of outage for all users is set to $p=0.1$.

\begin{figure}[h]
    \centering
    \includegraphics[width=0.45\textwidth]{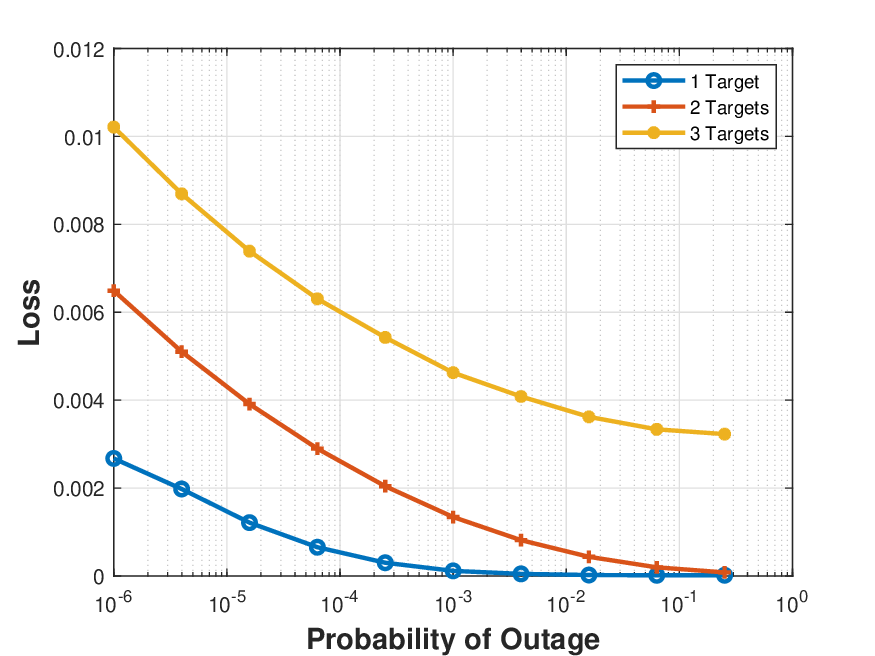}
    \caption{Loss versus probability of outage for different number of directions of interests (or targets)}
         
    \label{fig:Fig7}
\end{figure}

Fig. ~\ref{fig:Fig7} shows the difference between the loss for different number of directions of interests (or targets) and the loss of radar only system. We experimented with a system that includes two users and the outage SINR threshold is set to $\gamma_{th} = 5\mathrm{dB}$. Since the loss of radar-only system for different number of directions of interest is different, we used the difference to make fair comparison between the performance of different systems. As expected, the loss increases when we have more directions of interest.

\begin{figure}[ht]
    \centering
    \begin{subfigure}[b]{0.45\textwidth}
        \includegraphics[width=\textwidth]{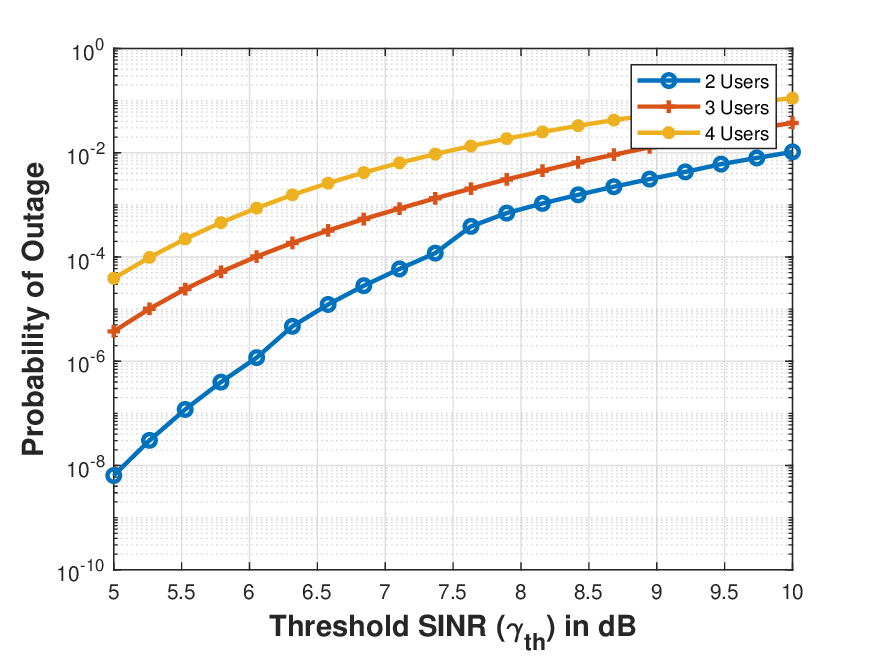}
        \caption{Probability of outage versus threshold SINR}
        \label{fig:5a}
    \end{subfigure}
    \begin{subfigure}[b]{0.45\textwidth}
        \includegraphics[width=\textwidth]{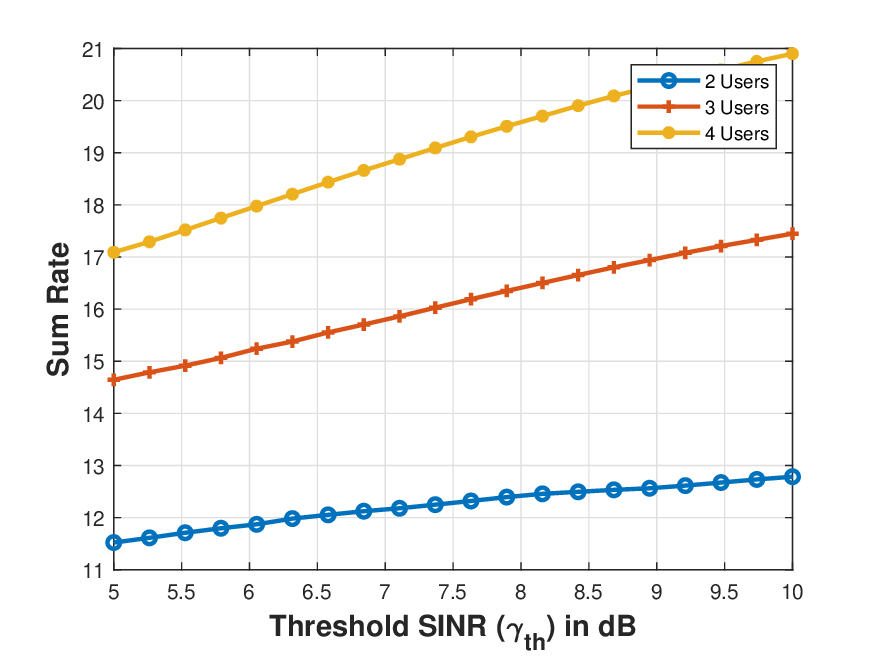}
        \caption{Sum rate (bps/Hz) versus different SINR thresholds, same setup as Fig.~\ref{fig:5a}}
        \label{fig: 5b}
    \end{subfigure}
    \caption{Performance of DFRC for the min-max probability of outage problem}
    \label{fig:Fig5}
\end{figure}

Fig. \ref{fig:Fig5} shows the performance of our proposed method to solve the min-max probability of outage problem. In our simulations, we set the MSE error threshold equal to $0.3$ and the cross correlation error threshold equal to $5$. Fig \ref{fig:5a} shows the maximum probability of outage among the users versus different values of outage threshold SINR. Fig \ref{fig: 5b} shows the sum rate of the same system. Comparing the results of this figure with that of Fig. \ref{fig:Fig3}, we see that this system is achieving higher sum rates. For example, with the case of 2 users and SINR threshold from 5 to 10, we see a change of sum rate from approximately 7 to 11 in Fig. \ref{fig:3a}, while in Fig. \ref{fig:5a} we see a change from 11 to 13. Another important observation is that the slope of change for the min-max system is lower. In fact, this system achieves high performance from very low SINR thresholds, therefore does not show a high rate of change as the SINR threshold changes.

\begin{figure}[h]
    \centering
    \includegraphics[width=0.45\textwidth]{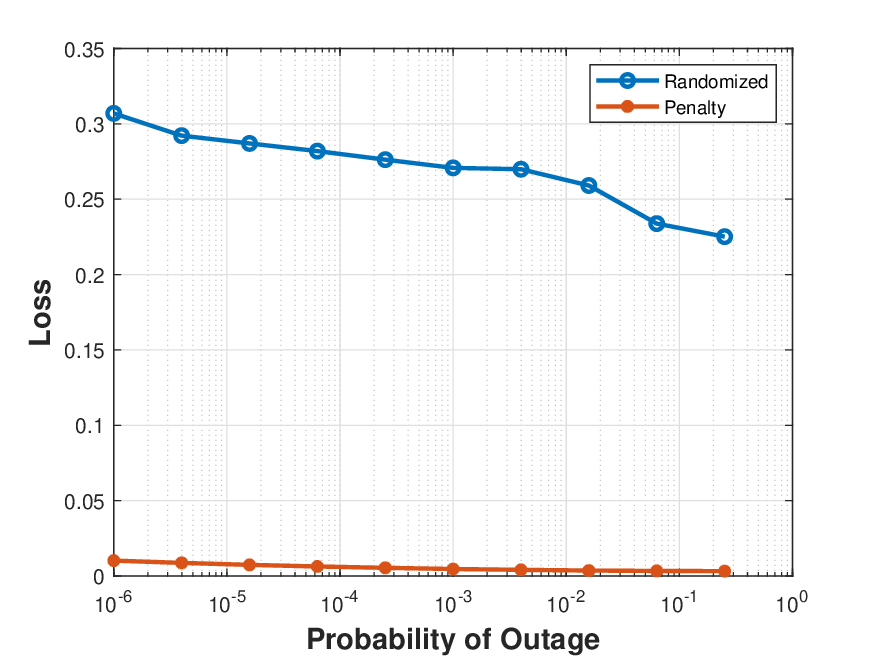}
    \caption{Loss versus probability of outage comparison for randomization method and the penalty method}
         
    \label{fig:Fig8}
\end{figure}

We also demonstrate the advantage of the penalty-based method over randomization. Using Gaussian randomization in the DFRC setting is challenging because we must ensure that the beamformers meet both the equal per-antenna power constraint and the probability of outage constraint. Here, we propose a heuristic approach, but it requires a very large number of randomized candidates to be effective in scenarios with more than two communication users. Let $\Tilde{\mathbf{W}}_1, \dots, \Tilde{\mathbf{W}}_K$ be the solutions of the optimization problem \eqref{opt:Nonrank1}, solved without rank-1 constraints. We draw one sample from the distributions $\mathcal{CN}(0, \Tilde{\mathbf{W}}_1), \dots, \mathcal{CN}(0, \Tilde{\mathbf{W}}_K)$ to get $\hat{\mathbf{w}}_1, \dots, \hat{\mathbf{w}}_K$. Then, we build a matrix $\hat{W}$ where the $k$-th column is equal to $\hat{\mathbf{w}}_k$ and normalize each row to have a norm equal to $P_T/N$. After that, we choose the columns of the normalized matrix as the rank-1 beamformers and check Constraint \eqref{constraint13b} (or equivalently Constraint \eqref{const:outage}) to make sure that the candidate beamformers satisfy the outage probability constraints. If the set of beamformers satisfy the constraints, we will store them, otherwise, we will disregard them. After having several sets of beamoformers, we check the objective function \eqref{loss} and select the set of beamformers that results in the smallest loss.  Fig. ~\ref{fig:Fig8} compares the performance of the DFRC when optimized by the proposed randomization method and by the proposed penalty method. We experimented with two users and three directions of interest. The outage SINR threshold was set to $\gamma_{th} = 5 \mathrm{dB}$. The optimal solution was chosen among the feasible solutions in 40,000 random drawings. As we can see, the penalty method outperforms the randomization method by a large margin. The cases with more users require very large number of randomized candidates and barely result in a feasible solution.

\begin{figure}[ht]
    \centering
    \begin{subfigure}[b]{0.45\textwidth}
        \includegraphics[width=\textwidth]{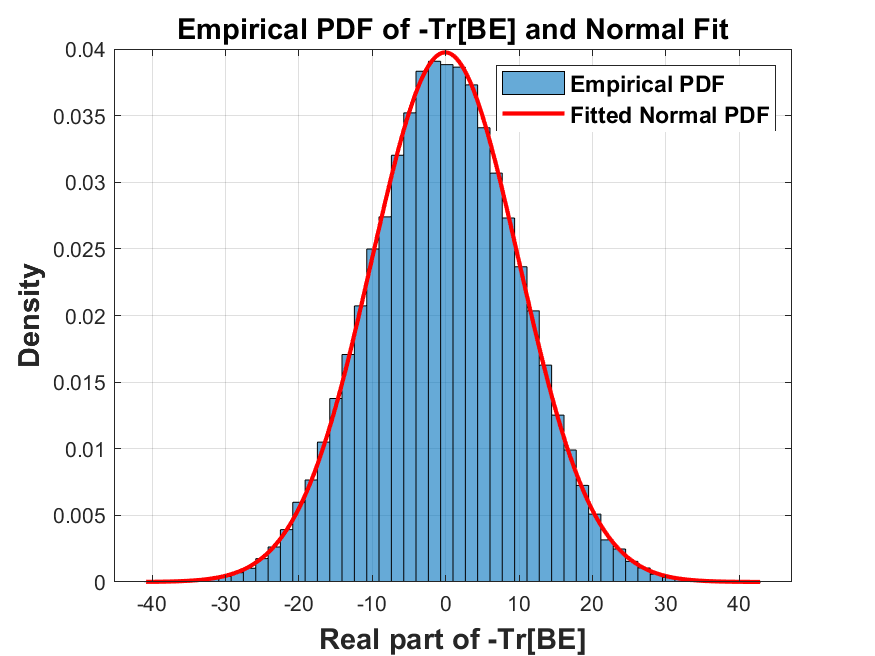}
        \caption{Uniform R.V.'s}
        \label{fig:9a}
    \end{subfigure}
    \begin{subfigure}[b]{0.45\textwidth}
        \includegraphics[width=\textwidth]{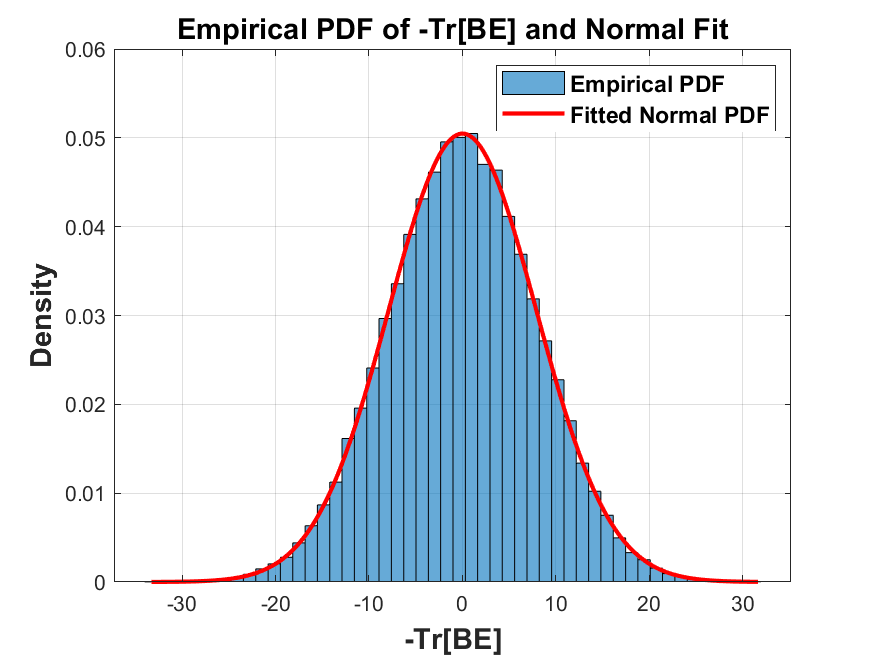}
        \caption{Non-uniform non-Gaussian R.V.'s}
        \label{fig: 9b}
    \end{subfigure}
    \caption{Validation of Gaussianity of $-\mathrm{Tr}[\mathbf{BE}]$ for dependent non-Gaussian elements in $\mathbf{E}$}
    \label{fig:Fig9}
\end{figure}

Finally, we proved that the random variables $-\mathrm{Tr}[\mathbf{B}_k \mathbf{E}_k]$ are asymptotically Gaussian. We provide simulation results to show that the result is also valid for small values of $N$. 
We consider a case with $N = 10$ ($10 \times 10$ matrices). 
Fig.~\ref{fig:Fig9} shows the histogram for 100,000 Monte Carlo simulations of dependent non-Gaussian random variables.
 We first create a correlation matrix $\mathbf{P}$, where $p_{ij} = \exp(-\lambda \lvert i - j \rvert)$. Using the Cholesky decomposition, we obtain $\mathbf{P} = \mathbf{L}\mathbf{L}^T$.
 Then, we create a vector $\mathbf{x}$ of length $N(N-1)/2$, containing independent complex normal random variables. The random vector $\mathbf{y} = \mathbf{L}\mathbf{x}$ will have a covariance matrix $\mathbf{P}$.
 While $\mathbf{y}$ is a normal random variable, we apply $G(\cdot)$, the CDF of the normal distribution, to the real and imaginary parts of $\mathbf{y}$ to transform them into uniform random variables. To make them zero-mean, we subtract $0.5 + j0.5$ from the uniform random variables.
 Note that the variables will be dependent, but the covariance matrix will not remain equal to $\mathbf{P}$. These R.V.'s constitute the upper triangle. The lower triangle is built by conjugating the upper triangle. The diagonal contains real independent uniform R.V's. Fig.~\ref{fig:9a} shows that the resulting random variable matches well to a Gaussian random variable.

To maintain the covariance $\mathbf{P}$, in another experiment, we let the real and imaginary parts of $\mathbf{x}$ be uniformly distributed in $[-0.5, 0.5]$. Following the same procedure, we obtain random variables $\mathbf{y} = \mathbf{L}\mathbf{x}$ with different distributions. Since $\mathbf{L}$ is a lower triangular matrix, each component of $\mathbf{y}$ is a sum of a number of uniform random variables. However, the correlation matrix of $\mathbf{y}$ will still be $\mathbf{P}$.
Fig.~\ref{fig: 9b} shows the simulation results for this setting which also shows a good match to a Gaussian random variable.

\begin{figure}[ht]
    \centering
    \begin{subfigure}[b]{0.45\textwidth}
        \includegraphics[width=\textwidth]{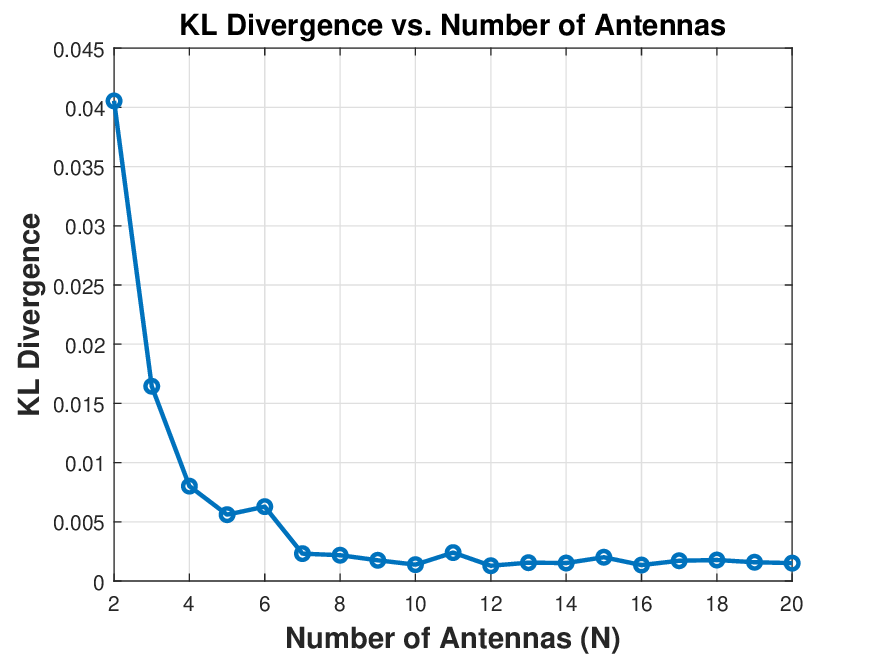}
        \caption{Uniform R.V.'s}
        \label{fig:10a}
    \end{subfigure}
    \begin{subfigure}[b]{0.45\textwidth}
        \includegraphics[width=\textwidth]{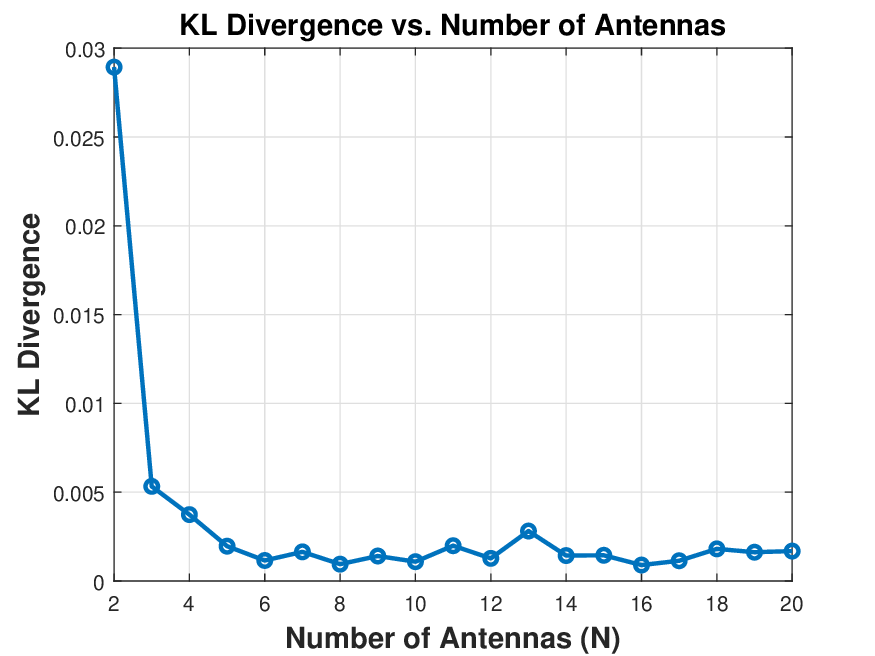}
        \caption{Non-uniform non-Gaussian R.V.'s}
        \label{fig: 10b}
    \end{subfigure}
    \caption{KL divergence between the resulting empirical distribution and the fitted Gaussian distribution}
    \label{fig:Fig10}
\end{figure}

To objectively demonstrate the Gaussian nature of the empirical distribution, we present the KL divergence between the fitted Gaussian distribution and the empirical distribution for various sizes of the random matrix $\mathbf{B}$ in Fig.~\ref{fig:Fig10}. 
As shown in Fig.~\ref{fig:Fig10}, the convergence is fast and the fitted Gaussian distribution is a valid representation for small values of $N$, for example $N=8$.

\section{Discussion and Concluding Remarks}






\label{conclusion}
We investigated transmit signal design for a dual-function radar and communication system. We considered the radar-centric and communication-centric scenarios. In the radar-centric scenario, we minimized the MSE between  the radar beampattern and a desired prespecified pattern in addition to minimizing the cross correlation between different targets of interest, while keeping the probability of outage for each communication user below a given value. In the communication-centric scenario, we minimized the maximum probability of outage among users, while satisfying a maximum beampattern mismatch and a maximum cross-correlation among different directions of interest. We have considered cross-correlation between different directions of interest which is not studied in related works that consider probability of outage for DFRC. We converted the corresponding optimization problem with probabilistic constraints into one with deterministic constraints and employed semidefinite relaxations to manage the problem's non-convex nature. We used a penalty method to find rank-1 solutions from the semidefinite programs. Numerical simulations verified the performance of the proposed methods. 
Previous studies in the literature typically assume that the elements of the covariance matrix error are independent Gaussian random variables. In our work, we extended this framework to accommodate more general distributions. A promising avenue for further research involves modeling the covariance matrix $\mathbf{C}$ as a random matrix and applying the tools of random matrix theory. Specifically, the covariance matrix can be represented as a Gaussian distribution on Riemannian symmetric spaces with non-positive curvature.
While these models are more complex and computationally demanding, they provide a more realistic characterization of the covariance matrix. Incorporating such models could lead to new optimization problems that require further investigation. Another extension is to consider the case where the outage events are correlated. These extensions would offer valuable insights and serve as meaningful future research directions. 

\appendices
\section{Gaussianity for Independent Entries}
\label{App:A}

For ease of notation, we introduce a new variable $Y_m$ which is defined as follows:
\begin{equation}
    Y_m =
\begin{cases}
b_{k,ii}e_{k,ii}, &m = i, \, i = 1, \dots, N,\\[5pt]
2\mathrm{Re}\{b_{k,ji}e_{k,ij}\}, &m = N + \sum_{p=1}^{j-i-1} (N - p) + i,\\ &\text{ for } 1\leq i < j\leq N.
\end{cases}
\end{equation}

Although the choice of indexing does not affect the results for the independent case, we choose the indexing used in Appendix \ref{App: B} for consistency. The variables are indexed diagonal by diagonal from the original matrix, i.e., the elements are arranged starting with the diagonal (where 
$i=j$), followed by elements in the upper triangle grouped by diagonals. 

Gaussianity of $-\mathrm{Tr}[\mathbf{B}_k \mathbf{E}_k]$ will be established based on the following theorem (Lyapunov CLT):
\begin{theorem}[\cite{schervish2014probability}]
    Suppose that the random variables \( X_1, X_2, \ldots \) are independent.
Let \( \mathbb{E}(X_i) = \mu_i \) and \( \text{Var}(X_i) = \sigma_i^2 \) for \( i = 1, \ldots, n \). Also, let

\[
Y_n = 
\frac{\sum_{i=1}^n X_i - \sum_{i=1}^n \mu_i}
{\left( \sum_{i=1}^n \sigma_i^2 \right)^{1/2}}.
\]
Assume that \( \mathbb{E}(|X_i - \mu_i|^3) < \infty \) for \( i = 1, 2, \ldots \) and

\[
\lim_{n \to \infty} 
\frac{\sum_{i=1}^n \mathbb{E}(|X_i - \mu_i|^3)}
{\left( \sum_{i=1}^n \sigma_i^2 \right)^{3/2}} = 0.
\]
Then, for each fixed number \( x \),

\[
\lim_{n \to \infty} \Pr(Y_n \leq x) = \Phi(x),
\]
where \( \Phi \) denotes the cumulative distribution function (c.d.f.) of the standard Gaussian distribution.

\end{theorem} 
For the independent case, we assumed that the random variables $e_{k,ij},  1\leq i\leq j\leq N$ are independent and the real and imaginary parts have equal variances. Using the theorem for random variables $Y_m$ and assuming that they satisfy the conditions of the theorem, we can conclude that $-\mathrm{Tr}[\mathbf{B}_k \mathbf{E}_k]$ is a Gaussian random variable. 
The theorem's conditions are not prohibitive. The first condition assumes that the third absolute moments are finite, as is the case for any practical scenario. 
The second condition ensures that the third absolute moments do not accumulate too fast compared to the second moments. 
These conditions are satisfied in almost all realistic scenarios.   
\section{Gaussianity for Dependent Entries}
\label{App: B}
There are multiple articles that present CLT for dependent random variables (e.g., \cite{rosenblatt1956central, ibragimov1962some}). We will use the results of \cite{herrndorf1985functional} which are less restrictive and do not assume stationarity of the sequence of random variables.

\begin{theorem}[\cite{herrndorf1985functional}]
    Let \( (X_j)_{j \in \mathbb{N}} \) be a sequence of random variables (R.V.'s) on some probability space \( (\Omega, \mathcal{F}, P) \). 
Assume that $\mathbb{E}[X_j] = 0$, $\mathbb{E}[X_j^2] < \infty$ for all $j \in \mathbb{N}$, and $S_n=\frac{1}{n} \sum_{j=1}^n X_j^2 \xrightarrow{P} \sigma^2$ for some $\sigma^2 > 0$, where $\xrightarrow{P}$ denotes convergence in probability. For any collection of R.V.'s \(\mathcal{X}\), let \(\sigma(\mathcal{X})\) denote the \(\sigma\)-algebra generated by \(\mathcal{X}\). For any two \(\sigma\)-algebras \(\mathcal{A}, \mathcal{B}\), let
\begin{equation}
    \label{alpha_mixing}
    \alpha(\mathcal{A}, \mathcal{B}) = \sup \left\{ \left| P(F \cap G) - P(F)P(G) \right| : F \in \mathcal{A}, G \in \mathcal{B} \right\}.
\end{equation}
The mixing coefficients \(\alpha(k)\) of the sequence \( (X_n)_{n \in \mathbb{N}} \) are defined as:
\[
\alpha(k) = \sup_{n \in \mathbb{N}} \alpha\left(\sigma(X_j : j \leq n), \sigma(X_j : j \geq n + k)\right), \quad k \in \mathbb{N}.
\]
If there exists \( \beta > 2 \) such that
\[
\sup_{j \in \mathbb{N}} \mathbb{E}[\lvert X_j \rvert^\beta] < \infty, \quad \text{and} \quad \sum_{k \in \mathbb{N}} \alpha(k)^{1 - 2/\beta} < \infty,
\]
where $\alpha(k)$ is defined according to \eqref{alpha_mixing}, then \( W_n \to W \), where for $n \in \mathbb{N}$, $W_n$ is defined by:
\[
W_n(t, \omega) = \frac{S_{\lfloor nt \rfloor}(\omega)}{\sigma\sqrt{n}}, \quad t \in [0, 1], \, \omega \in \Omega,
\]
and the notation $W_n \xrightarrow{d} W$ denotes the weak convergence of the distribution of $W_n$ to the standard Wiener measure $W$.
\label{thm:2}
\end{theorem}
Applying Theorem \ref{thm:2} to random variables $Y_m$ and assuming that they satisfy the conditions of the theorem, we conclude that $-\mathrm{Tr}[\mathbf{B}_k \mathbf{E}_k]$ is a Gaussian random variable. As discussed in the main body of the paper, the conditions (bounded variance, at least one larger than 2 bounded moment, and decaying correlation) are not prohibitive and are easily satisfied in most cases. Also, as validated by the simulations in Fig. \ref{fig:Fig9}, even in the case of a small number of antennas, the distribution converges to a Gaussian distribution.

 \bibliography{mybibliography}
 \bibliographystyle{IEEEtran}

\end{document}